\documentclass[12pt,english]{article}
\usepackage[T1]{fontenc}
\usepackage[latin9]{inputenc}
\usepackage[letterpaper]{geometry}
\usepackage{float}
\usepackage{amsmath}
\usepackage{graphicx}
\usepackage{setspace}
\usepackage{amssymb}
\usepackage{babel}

\begin{document}

\title{Hawking radiation and thermodynamics of dynamical black holes 
in phantom dominated universe}

\author{Khireddine Nouicer%
\thanks{E-mail:khnouicer@univ-jijel.dz%
}}

\date{}
\maketitle
\begin{quote}
\begin{center}
Laboratory of Theoretical Physics (LPTh) and Department of Physics,
Faculty of Sciences, University of Jijel, Bp 98 Ouled Aissa, Jijel
18000, Algeria 
\par\end{center}\end{quote}

\begin{abstract}
The thermodynamic properties of dark energy-dominated universe in
the presence of a black hole are investigated in the general case
of a varying equation-of-state-parameter $w(a)$. We show that all
the thermodynamics quantities are regular at the phantom divide crossing,
and particularly the temperature and the entropy of the dark fluid
are always positive definite. We also study the accretion process
of a phantom fluid by black holes and the conditions required for
the validity of the generalized second law of thermodynamics. As a
results we obtain a strictly negative chemical potential and an equation-of-state
parameter $w<-5/3.$

\end{abstract}

\section{Introduction}

The discovery that the universe is currently undergoing a period of
accelerating expansion, obtained from the observations of type Ia
supernovae \cite{Perlmutter-1,Riess}, inaugurate an exciting era
of intense theoretical research. A variety of possible solutions to
understand the mechanism driving this accelerating expansion have
been debated during this decade including the cosmological constant,
exotic matter and energy, modified gravity, anthropic arguments, etc.
The most favored ones are dark energy based models and modified gravity
theories such $f(R)$ gravity \cite{Capozziello}, and Dvali-Gabadadze-Porrati
model of gravity \cite{DGP} with an equation of state (EoS) parameter
$w=P/\varrho<-1$ (where $\rho$ and $P$ are the energy density and
pressure of the cosmic fluid, respectively). The dark energy is frequently
modeled as an homogeneous scalar field, and according to the value
of the EoS parameter $w$, three different cases can be distinguished:
\textit{quintessence scalar fields} ($-1<w<-1/3$)
with a positive kinetic term ; \textit{cosmological constant} ($w=1$),
where only the potential term contributes to both the pressure and
the energy density of the field; finally, scalar fields with a negative
kinetic term dubbed \textit{phantom fields} ($w<-1$).
In the latter case the universe will suffer a crucial fate where the
energy density and the scale factor diverge in finite time, ripping
apart all bound systems of the universe, before the universe approaches
the so-called Big-Rip singularity \cite{Caldwell,McIness}. It is
also well known that in phantom fluid models with big rip singularity,
quantum gravity effects become dominant in the neighboring of the
big rip time \cite{Elizalde,Nojiri1,Nojiri2,Nojiri3}. However, even
phantom fluid based models suffer from quantum instabilities \cite{Cline}
and violation of the strong and dominant classical energy conditions
\cite{Schulz}, the phantom fluids are favored by the cosmic microwave
background experiments combined with large scale structure data, the
Hubble parameter measurement and luminosity measurements of Type Ia
supernovae \cite{Melchiorri}. 

An other important and growing field currently under investigation
is related to the thermodynamic properties of an expanding universe
\cite{Thermo}. Recent studies on phantom thermodynamics show that
the entropy of the universe is negative \cite{Brevik}, while the
generalized second law of gravitational thermodynamics (GSL) is satisfied,
$\dot{S}_{f}+\dot{S}_{C}\geq0$, where $S_{f}$ is the phantom fluid
entropy and $S_{C}$ is the entropy of the cosmological horizon \cite{Izquierdo}.
An other point under debate is the influence of a non-zero chemical
potential on the phantom thermodynamics and its relation with the
GSL \cite{Lima-Pereira,Horvath}.

In this paper the thermodynamic properties of black holes immersed
in dark energy-dominated expanding universe and the accretion process
of phantom energy onto black holes are investigated. The first paper
dealing with the later process is due to Babishev et al \cite{Babichev},
where ignoring the backreaction effect of the phantom fluid on the
black hole, they found that the change rate of the black hole mass
is negative. However, in recent scenarios where the backreaction is
taken into account \cite{Faraoni1,Faraoni2}, it is found that the
mass of the black hole is always an increasing function in an expanding
Friedman-Robertson-Walker (FRW) universe.

The organization of this paper is as follows: in Section 2 we review
the exact solution recently obtained in \cite{Faraoni1,Faraoni2}
, and describing a black hole embedded in an expanding FRW universe.
In section 3, we examine the Hawking radiation at the apparent horizon,
and compare with magnitude of the phantom energy accretion process.
We will show that the former is highly suppressed, particularly at
late times. In section 4, we study in an unified and general way the
thermodynamics of cosmological black hole embedded in an expanding
(FRW) universe with a general EoS parameter $w\left(a\right)$, and
particularly we obtain solutions realizing the crossing of the phantom
divide line. In section 5, the stability of the solutions of section
4 under the quantum correction due to the conformal anomaly is established.
In section 6, we study the conditions required for the validity of
the GSL when the black hole is immersed in phantom energy-dominated
FRW universe. Particularly, in order to protect the GSL, we obtain
a critical mass of the black hole of the order of the solar mass for
particular values of the parameter $\alpha_{0}=-\mu_{0}n_{0}/\rho_{0},$
where $\mu_{0},\: n_{0},\:\rho_{0}$ are the present day values of
the chemical potential, the particle density and the energy density,
respectively. Finally, we discuss and summarize our results in section
7.

\section{Cosmological expanding black hole}

The first solution of Einstein's theory of general relativity describing
a black hole like object embedded in an expanding universe was introduced
by McVittie in 1933 \cite{McVittie}, and is given in isotropic coordinates
by

\begin{equation}
ds^{2}=-\frac{\left(1-\frac{M_{0}}{2a(t)r}\right)^{2}}{\left(1+\frac{M_{0}}{2a(t)r}\right)^{2}}dt^{2}+a^{2}(t)\left(1+\frac{M_{0}}{2a(t)r}\right)^{4}\left(dr^{2}+r^{2}d\Omega^{2}\right),\label{eq:McVittie}\end{equation}
where $a(t)$ is the scale factor and $M_{0}$ is the mass of the
black hole in the static case. In fact, when $a(t)=1,$ it reduces
to the Schwarzschild solution. When the mass parameter is zero, the
McVittie reduces to a spatially flat FRW solution with the scale factor
$a(t)$. The global structure of (\ref{eq:McVittie}) has been studied
and particularly it has been shown that the solution possesses a spacelike
singularity on the 2-sphere $r=M_{0}/2,$ and cannot describe an embedded
black hole in an expanding spatially flat FLRW universe \cite{Sussman,Nolan}.
On the other hand the McVittie solution is constrained by the non-accretion
condition onto the central mass, and therefore is not suitable for
a study of cosmic fluid accretion process onto black holes embedded
in an expanding FRW universe. 

In the following we adopt the new solution describing a black hole
embedded in a spatially flat FRW universe\cite{Faraoni1,Faraoni2}

\begin{equation}
ds^{2}=-\frac{B^{2}(r)}{A^{2}(r)}dt^{2}+a^{2}(t)A^{4}(r)\left(dr^{2}+r^{2}d\Omega^{2}\right),\label{eq:metric01}\end{equation}
where $A(r)=\left(1+\frac{GM_{0}}{2r}\right)\textrm{, }B(r)=\left(1-\frac{GM_{0}}{2r}\right)$,
$a(t)$ is the scale factor and $M_{0}$ is the mass of the black
hole in the static case. In fact, when $a(t)=1,$ the solution (\ref{eq:metric01})
reduces to the Schwarzschild solution, while when the mass parameter
is zero, it reduces to a spatially flat FRW solution with the scale
factor $a(t)$.

Using the areal radius $\widetilde{r}=r\left(1+\frac{GM_{0}}{2r}\right)^{2}\,\textrm{and}\: R=a\widetilde{r},$
the metric takes the following suitable Painlev�-Gullstrand form

\begin{flalign}
ds^{2}= & -\left[\left(1-\frac{2GM_{0}a}{R}\right)-\frac{R^{2}H^{2}}{\left(1-\frac{2GM_{0}a}{R}\right)}\right]dt^{2}+\left(1-\frac{2GM_{0}a}{R}\right)^{-1}dR^{2}\label{eq:metric02}\\
 & -2RH\left(1-\frac{2GM_{0}a}{R}\right)^{-1}dtdR+R^{2}d\Omega^{2},\nonumber \end{flalign}
 where $H=\dot{a}/a$ is the Hubble parameter and over dot stands
for derivative with respect to the cosmic time. The term $R^{2}H^{2}$
plays the role of variable cosmological constant. Now, we introduce
the time transformation $t\longrightarrow\bar{t}$ to remove the $dtdR$
term

\begin{equation}
d\overline{t}=F^{-1}\left(t,R\right)\left[dt+\frac{HR}{\left(1-\frac{2GM_{0}a}{R}\right)^{2}-H^{2}R^{2}}dR\right],\label{eq:Time-Transform}\end{equation}
 where the integrating factor $F\left(t,R\right)$ satisfy

\begin{equation}
\partial_{R}F^{-1}=\partial_{t}\left[\frac{F^{-1}HR}{\left(1-\frac{2GM_{0}a}{R}\right)^{2}-H^{2}R^{2}}\right].\label{eq:diff-exact}\end{equation}
 Substituting $\left(\ref{eq:Time-Transform}\right)$ into $\left(\ref{eq:metric02}\right)$
and replacing $\overline{t}\longrightarrow t$ , we obtain\begin{flalign}
ds^{2} & =-\left[\left(1-\frac{2GM_{0}a}{R}\right)-\frac{R^{2}H^{2}}{\left(1-\frac{2GM_{0}a}{R}\right)}\right]F^{2}dt^{2}\label{eq:metric03}\\
 & +\left[\left(1-\frac{2GM_{0}a}{R}\right)-\frac{R^{2}H^{2}}{\left(1-\frac{2GM_{0}a}{R}\right)}\right]^{-1}dR^{2}+R^{2}d\Omega^{2}.\nonumber \end{flalign}
 The apparent horizons (AH) are solutions of $h^{ab}\partial_{a}R\partial_{b}R=0,$
which leads to

\begin{equation}
\left(1-\frac{2Gm_{H}(t)}{R}\mp RH\right)\left|_{R_{A}}\right.=0,\label{eq:AHequation}\end{equation}
 where we introduced the Hawking-Hayward quasi-local mass \begin{equation}
m_{H}(t)=M_{0}a(t).\end{equation}
A remarkable feature of this quantity is that it is coordinate-independent,
and consequently is recognized as the physically relevant mass of
the black hole. Obviously, it is always increasing in an expanding
universe \cite{Faraoni2}. Therefore, the calculation of the change
rate of the black hole mass will lead to opposite conclusions to that
of Babishev et al. \cite{Babichev}.

Discarding the unphysical branch with the lower sign in \ref{eq:AHequation},
the AH are given by

\begin{flalign}
R_{B}= & \frac{1}{2H}\left(1-\sqrt{1-8G\dot{m}_{H}(t)}\right),\quad R_{C}=\frac{1}{2H}\left(1+\sqrt{1-8G\dot{m}_{H}(t)}\right),\label{eq:AHB}\end{flalign}
where $R_{C}$ and $R_{B}$ are the cosmological and the black hole
AH, respectively. Note that the AH coincide at a time $t_{*}$ defined
by $\dot{a}(t_{*})=1/8GM_{0}.$ This coincidence takes place in a
future or past universe depending on the kind of fluid accretion onto
the black hole. 

Let us now consider that the fluid is described by a imperfect fluid
with the stress-energy tensor given by

\begin{equation}
T_{\mu\nu}=\left(P+\rho\right)u_{\mu}u_{\nu}+Pg_{\mu\nu}+q_{\mu}u_{\nu}+q_{\nu}u_{\mu},\end{equation}
where $u^{\mu}=\left(\frac{A}{B},0,0,0\right)$ is the fluid four
velocity and $q^{\mu}=\left(0,q,0,0\right)$ is a spatial vector field
describing the radial energy current. Written in terms of the comoving
AH, the solutions of the Einstein equations of motion are \cite{Faraoni2},

\begin{flalign}
H= & \frac{8\pi G}{3}R_{A}\rho,\quad3H+\frac{\dot{2H}}{H}=-8\pi GR_{A}p.\label{eq:E3}\end{flalign}
Assuming a radial heat inflow ($q<0$), the change rate of the black
hole mass is then\begin{equation}
\dot{m}_{H}=GaB^{2}\mathcal{A}\left|q\right|,\label{eq:rate}\end{equation}
where $\mathcal{A=\int\int}d\theta d\varphi\sqrt{g_{\Sigma}}=4\pi r^{2}a^{2}A^{4}$
. This relation shows that the Hawking-Hayward quasi-local mass is
always increasing.

\section{Hawking radiation of apparent horizon}

For the apparent horizon to exist in an expanding universe we have
the following condition 

\begin{equation}
H\left(t\right)\leq\frac{1}{8Gm_{H}\left(t_{*}\right)},\label{eq:Hcond}\end{equation}
This condition requires us to consider massive objects for which $Gm_{H}H$
is a small quantity.  Using the present day value of the Hubble parameter,
we estimate the critical BH mass as $m_{H}(t_{*})\lesssim10^{23}\, M_{\odot}$.
This is the condition for which the engulfing of the universe by the
BH is prevented \cite{Martin}. Using the areal radius $R=ar\left(1+\frac{GM_{0}}{2r}\right)^{2}$,
and solving for $r$ , we obtain one physical solution given by\begin{equation}
r_{A}=\frac{1}{2a}\left(R_{A}-Gm_{H}+\sqrt{R_{A}^{2}-2R_{A}Gm_{H}}\right).\end{equation}
For the solution to be real, we have to impose the condition\begin{equation}
R_{A}>2Gm_{H},\label{eq:GmH}\end{equation}
which is true only in an expanding universe due to the relation $R_{A}\left(1-HR_{A}\right)=2Gm_{H}$.
We note that the equality sign has been discarded in (\ref{eq:GmH}),
since it leads to $HR_{A}=0$, which does not hold for $m_{H}\neq0.$
The relation (\ref{eq:GmH}) means that we are dealing with systems
which are small compared to the cosmological curvature, and that distances
below $2l_{pl}$ are naturally excluded.

Assuming now the EoS, $p=w\rho,$ and substituting (\ref{eq:GmH})
in Friedmann equations, one finds the following upper bounds on the
densities

\begin{equation}
\rho\left(t\right)<\frac{3m_{pl}^{6}}{128\pi m_{H}^{2}\left(t\right)},\quad\left|p\right|<\frac{3\left|w\right|m_{pl}^{6}}{128\pi m_{H}^{2}\left(t\right)}.\label{eq:energy-pressure}\end{equation}
Therefore, in case of phantom energy driven expansion, the finite
increase of densities with time, will avoid the big rip singularity.
 On the other, since the BH mass increases with time, the upper bounds
(\ref{eq:energy-pressure}), become very small. As an estimate of
the energy density we have $\rho\left(t\right)<0.02\times10^{-10}\,\left(\textrm{GeV}\right)^{4},$
for the smallest super-massive BH detected in the dwarf Seyfert 1
galaxy POX 52 with $m_{H}\sim10^{5}\, M_{\odot}$ \cite{Rees}. On
the other hand if $m_{H}\left(t_{*}\right)\sim M_{\odot}$, we have
$\rho\left(t\right)<0.022\,\left(\textrm{GeV}\right)^{4}.$ Even if
we take $m_{H}$ of the order of the Planck mass we have $\rho\left(t\right)<\frac{3m_{pl}^{4}}{128\pi}.$ 

We now consider the Hawking radiation from the cosmological AH. The
temperature on the AH is defined by $T_{A}=\left|\kappa_{A}\right|/2\pi$,
where $\kappa_{A}$ is the surface gravity. In dynamical spacetime
there is no timelike Killing vector, and the usual definition of the
surface gravity may be modified. In this case case the surface gravity
is related to the so called trapping horizon. Here we follow the work
of S. A. Hayward \cite{Hayward}, where the surface gravity is defined
by $K^{b}\nabla_{\left[a\right.}K_{\left.b\right]}=\kappa K_{a}$,
$\left(a,b=0,1\right)$, with $K^{a}=-\varepsilon^{ab}\nabla_{b}R$
the Kodama vector corresponding to the background given by Eq.(\ref{eq:metric02}).
Evaluating all the quantities on the trapping horizon, we get the
simplified form, $\kappa_{A}=\frac{1}{2}\square_{h}R$, where the
metric $h_{ab}$ is defined by $ds^{2}=h_{ab}dx^{a}dx^{b}+R(x)d\Omega^{2}$.
Performing the calculation, one finally finds\begin{equation}
\kappa_{A}=\frac{m_{H}}{R_{A}^{2}}-H-\frac{\dot{H}}{2H}.\end{equation}
Writing $m_{H}$ is terms of the AH, the temperature at the AH becomes

\begin{equation}
T_{A}=\frac{1}{4\pi R_{A}}\left|1-\frac{R_{A}}{H}\left(3H^{2}+\dot{H}\right)\right|,\label{eq:Temperature-1}\end{equation}
Expanding (\ref{eq:Temperature-1}) to first order in $Gm_{H}$ at
the black hole AH and cosmic AH, we obtain \begin{equation}
T_{C}=\frac{H}{2\pi}\left|1+\frac{\dot{H}}{2H^{2}}\right|-\frac{G\dot{m}_{H}}{2},\label{eq:TCapprox}\end{equation}
\begin{equation}
T_{B}=\frac{1}{8\pi Gm_{H}}-\frac{GH^{2}m_{H}}{2\pi}-\frac{4H^{2}+\dot{H}}{4\pi H},\label{eq:TBapprox}\end{equation}
respectively. Using (\ref{eq:GmH}) in the first factor, along with
$R_{C}\leq1/H$ in the second factor, and $\frac{\dot{H}}{H^{2}}=\frac{3}{2}(1+w)$,
one obtains an instantaneous upper bound for temperature at cosmic
AH\begin{equation}
T_{C}<\frac{\left|1-3w\right|}{16\pi}\frac{m_{pl}^{2}}{m_{H}(t)}.\label{eq:Tmax}\end{equation}
The relation (\ref{eq:TCapprox}) shows that the accretion process
tends to lower the Hawking temperature at the cosmic AH. We note that
in (\ref{eq:TBapprox}) we have not used the absolute value, as a
consequence of the equivalence principle \cite{Arraut}. A similar
behavior is also observed if one anticipates and use the scale factor
(\ref{eq:scalefactor}) derived in section 4, for a phantom dominated
era. In fact, one can see an unusual behavior of the temperature when
approaching the time $t_{*}.$ As it is shown in Fig. 1, the temperature
at the cosmic AH increases with time, reaches a maximum, then begins
to decrease, and stops at the end time $t_{*}$. This means that the
Hawking radiation at the cosmological AH will decrease rapidly in
favor of a hugh increasing of the accretion of phantom fluid. On the
hand, the temperature of the black hole AH starts to fall at early
stage because of the increasing of the black hole mass by accretion
of phantom energy, and at late stage begins incraesing. However, the
latter strange behavior is mainly due to the absolute value in the
definition of temperature. 

An independent derivation of (\ref{eq:energy-pressure}) and (\ref{eq:Tmax})
can be performed by using the following simple arguments. Imposing
positivity of the $\left(-g_{tt}\right)$ component of the metric,
and replacing the quasi-local mass by a density energy we have, $0<-g_{tt}=1-2Gm_{H}/R-\frac{R^{2}H^{2}}{1-2Gm_{H}/R}=1-\left(8\pi G/3\right)\rho R^{2}-\frac{R^{2}H^{2}}{1-\left(8\pi G/3\right)\rho R^{2}}$.
Hence, $\rho<\left(1-RH\right)/$$\left(8\pi GR^{2}/3\right)$. Using
$R>2Gm_{H},$ one obtains $\rho<3m_{pl}^{6}/\left(32\pi m_{H}^{2}\right)$,
which is consistent with (\ref{eq:energy-pressure}). If one uses
the Stefan-Boltzmann law $\rho=\sigma T^{4},$ and the quantum mechanical
relation $R>1/T$ \cite{Massa}, one finds $T<\left(3/4\pi\sigma\right)m_{pl},$
which of the same order of magnitude as (\ref{eq:Tmax}) for $m_{H}=m_{pl}.$
Finally, we point that the extremal case corresponding to $T_{C}=0$
never occurs, since in this case we must have $R_{C}=\frac{1}{3H+\dot{H}/H^{2}},$
which contradicts the condition $R_{C}\geq1/\left(2H\right).$

\begin{figure}[H]
\begin{centering}
\includegraphics[width=8cm,height=8cm]{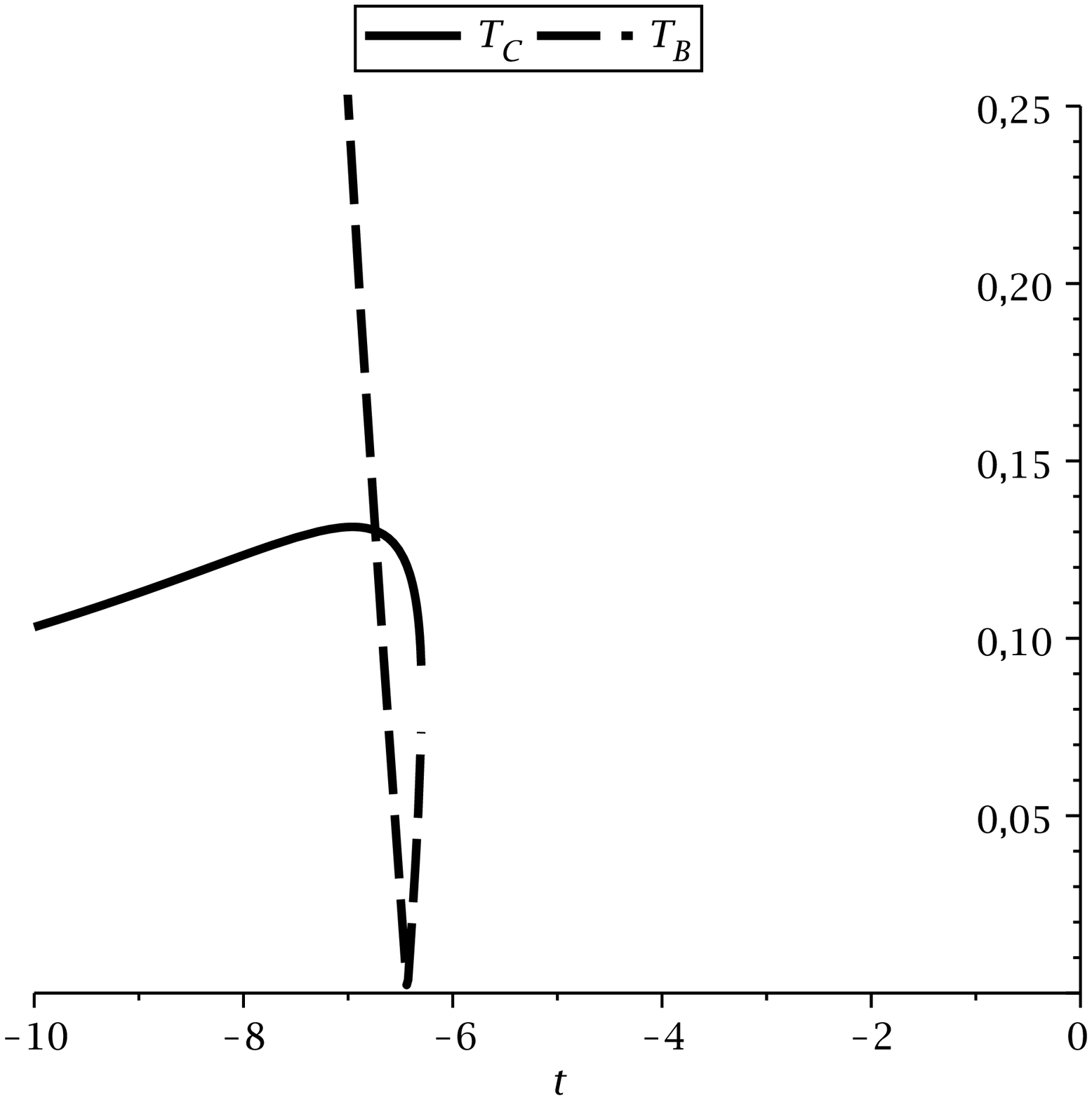}
\par\end{centering}

Figure 1: Variation of temperature associated with the black hole
and cosmic AH with time for $M_{0}=0.25$.

\end{figure}

Now, neglecting the accretion of radiation in phantom energy dominated
era, and taking into account only the semi-classical Hawking evaporation
and the phantom energy accretion term, the differential equation for
the black hole mass reads
\begin{equation}
\frac{dm_{H}}{dt}=-4\pi R_{B}^{2}\sigma T_{B}^{4}+m_{H}H,
\end{equation}
where $\sigma=N\pi^{2}/120$ is the Stefan-Boltzmann constant for
massless fields with effective degree of freedom $N.$ Substituting
the Hawking temperature associated with the BH apparent horizon, we
obtain
\begin{equation}
\frac{dm_{H}}{dt}=-\frac{\sigma H^{2}\left|1-\frac{3}{4}\left(1-\sqrt{1-8Gm_{H}H}
\right)\left(1-w\right)\right|^{4}}{16\pi^{3}\left(1-\sqrt{1-8Gm_{H}H}\right)^{2}}+m_{H}H.
\end{equation}
This is a complicated relation for $m_{H}$, whose behavior is shown
in fig. 2. Since the two terms are in competition, there exist a transition
time, the \textit{phantom time}, after which the accretion process
dominates and the BH mass increases. Consequently, the BH does not
lose but gain mass due to hugh accretion of dark energy. Indeed, it
easy to show that the maximal rate gain mass for $m_{H}\neq0,$ is
$\left(dm_{H}/dt\right)_{max}\sim m_{pl}^{2}.$ It is important to
note that the accretion term becomes predominant at earlier times
for massive BH. 
Next, we perform the same analysis on the variation of the mass inside
the cosmic AH. Let us define the ratio between the radiation and the
accretion term \begin{equation}
\eta_{C}\left(t\right)=\left|\frac{\dot{m}_{\textrm{Haw}}}{\dot{m}_{\textrm{ph}}}\right|=\frac{4\pi R_{C}^{2}\sigma T_{C}^{4}}{m_{H}H}.\label{eq:eta}\end{equation}
Substituting (\ref{eq:Temperature-1}), and repeating the procedure
leading to (\ref{eq:Tmax}), one finds\begin{equation}
\eta_{C}\left(t\right)<\frac{N\left|1-3w\right|^{4}}{245760\pi}\left(\frac{m_{pl}}{m_{H}}\right)^{2}.\label{eq:eta-1}\end{equation}
This result clearly shows that the Hawking radiation at the cosmic
AH is insignificant, even near the Planck scale. 

\begin{figure}[H]

\includegraphics[width=6cm,height=8cm]{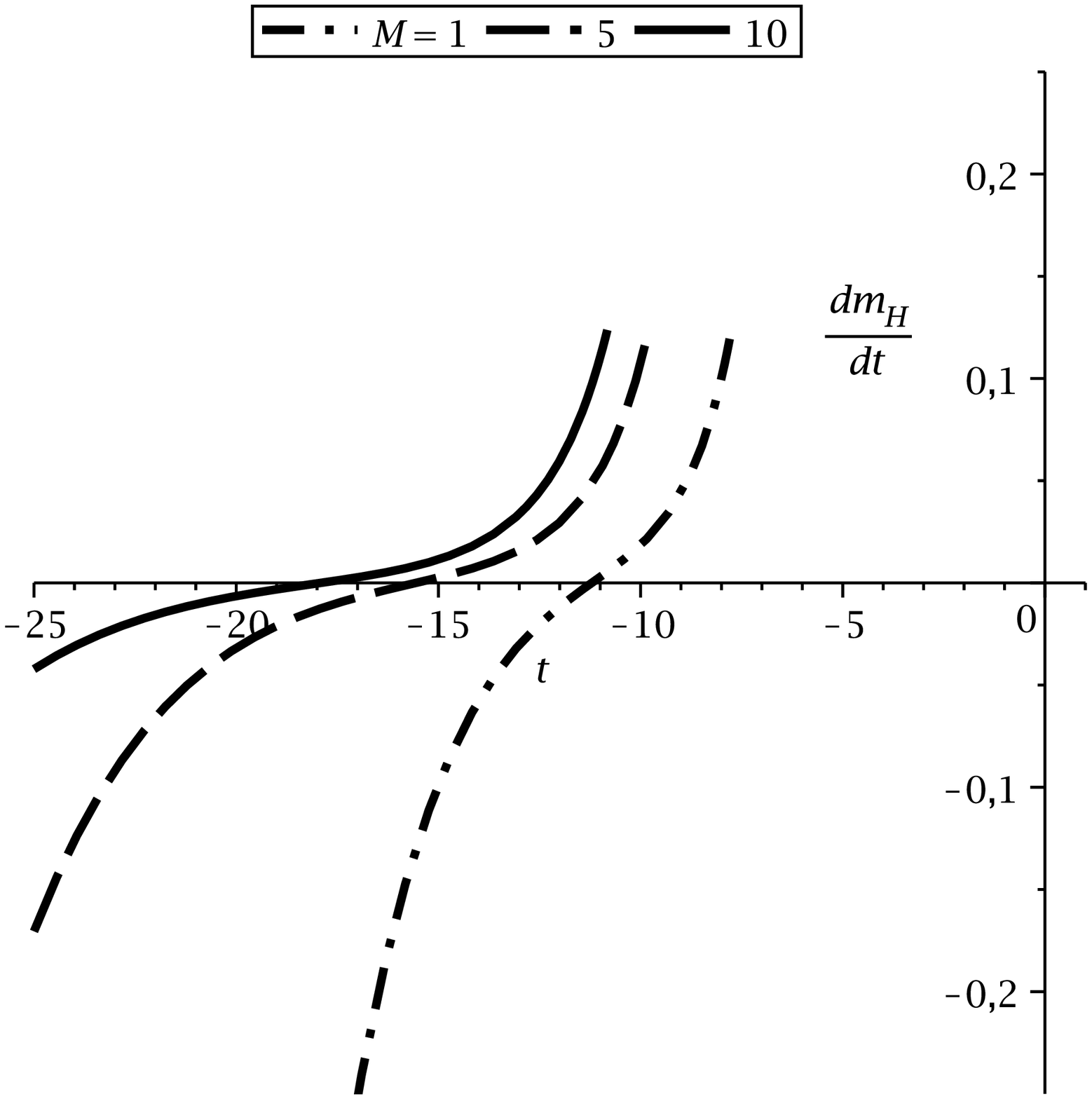}\includegraphics[width=6cm,height=8cm]{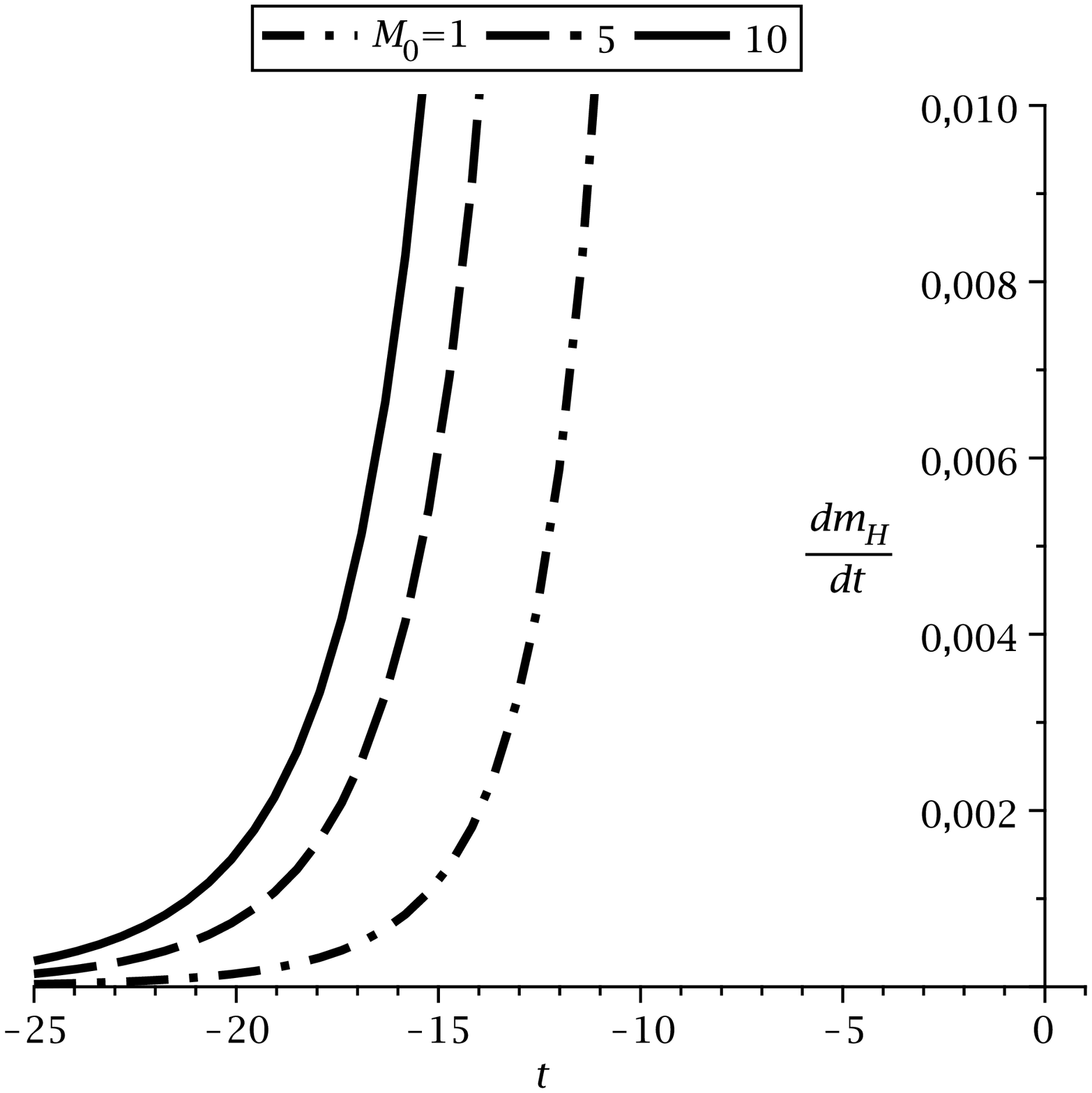}

Figure 2: Variation of total mass associated with the black hole AH
(left panel) and cosmic AH (right panel) with time.

\end{figure}

Finally, let us point an other crucial feature of the model of dynamical
black hole considered in this paper. If the expansion of the universe
is driven by phantom fluid, the black hole AH expands while the cosmic
AH shrinks as the universe expand until they meet at $R_{crit}=1/(2H)$
at time $t_{*}$ solution of (\ref{eq:Hcond}). At times $t>t_{*},$
both the AH disappear living a proper singularity, well before the
big-rip singularity is reached. The question if this singularity is
naked or located inside the AH, and its connection with the violation
of the Cosmic Censorship Conjecture (CCH), is still under debates
\cite{Faraoni3,Sun}. To avoid discussing this topics, which are out
of the scope of the present paper, and the fact that the phantom driven
expansion of the universe is non singular, and that the radiation
power can be neglected in comparison to the accretion of phantom fluid
onto the BH, particularly at later stage, we limit the analysis in
the remaining sections to the interval $t\leq t_{*}$, far from the
big rip singularity, and then purely classical treatment will be considered.

\section{Thermodynamics with varying $w$}

We now consider the thermodynamics properties of the solution described
in section 2, with a variable EoS parameter, $p\left(a\right)=w\left(a\right)\rho\left(a\right)$.
The particle fluid and entropy fluid currents, $N^{\mu}$ and $S^{\mu}$
are given by

\[
N^{\mu}=nu^{\mu},\: S^{\mu}=su^{\mu},\]
where $n$ and $s$ are the densities of particle number and entropy,
respectively. The conservations laws, $T_{\;\;;\nu}^{\mu\nu}=0,$
$N_{\;;\mu}^{\mu}=0$ and $S_{\;;\mu}^{\mu}=0,$ computed on the background
given by (\ref{eq:metric01}), give the following set of differential
equations

\begin{equation}
\dot{\rho}+\frac{\dot{R}_{A}}{R_{A}}\rho+\frac{3}{2}H\left(\rho+p\right)=0,\label{eq:continuity}\end{equation}

\begin{equation}
\dot{n}+\left(\frac{\rho}{\rho+p}\frac{\dot{R}_{A}}{R_{A}}+\frac{3}{2}H\right)n=0,\end{equation}
 \begin{equation}
\dot{s}+\left(\frac{\rho}{\rho+p}\frac{\dot{R}_{A}}{R_{A}}+\frac{3}{2}H\right)s=0.\end{equation}
The solutions of the above equations are

\begin{equation}
\rho\left(a\right)=\rho_{0}\left[\frac{R_{A}\left(a_{0}\right)}{R_{A}\left(a\right)}\right]\left[\frac{a_{0}^{\frac{3}{2}\left(1+w_{0}\right)}}{a^{\frac{3}{2}\left(1+w\left(a\right)\right)}}\right]\exp\left[\frac{3}{2}\int_{a_{0}}^{a}daw'\left(a\right)\ln a\right],\label{eq:rho}\end{equation}

\begin{equation}
n\left(a\right)=n_{0}\left(\frac{a_{0}}{a}\right)^{\frac{3}{2}}\exp\left[\int_{a}^{a_{0}}\frac{R'(a)}{\left(1+w(a)\right)R(a)}da\right],\label{eq:n}\end{equation}

\begin{equation}
s\left(a\right)=s_{0}\left(\frac{a_{0}}{a}\right)^{\frac{3}{2}}\exp\left[\int_{a}^{a_{0}}\frac{R'(a)}{\left(1+w(a)\right)R(a)}da\right],\label{eq:s}\end{equation}
where the prime stands for derivative with respect to the scale factor,
and $\rho_{0},$ $n_{0}$, $s_{0}$ are the present day values of
the corresponding quantities assumed to be positive definite. Here
we note the important corrections due to the presence of the black
hole. 

In the pure dark energy-dominated universe, $M_{0}=0,$ we have $R_{B}=0$
and $R_{C}=1/H$. Consequently, using the equations of motion (\ref{eq:E3}),
we show that

\begin{equation}
\frac{R'_{C}\left(a\right)}{R_{C}\left(a\right)}=\frac{3}{2}\left(1+w\left(a\right)\right)H\left(a\right).\label{eq:pure}\end{equation}
Substituting in Eqs.(\ref{eq:rho}-\ref{eq:s}) we obtain \cite{Saridakis}

\begin{flalign}
\rho\left(a\right)=\rho_{0}\left[\frac{a_{0}^{3\left(1+w_{0}\right)}}{a^{3\left(1+w\left(a\right)\right)}}\right]\exp\left[3\int_{a_{0}}^{a}daw'\left(a\right)\ln a\right],\quad n\left(a\right)= & n_{0}\left(\frac{a_{0}}{a}\right)^{3},\quad s\left(a\right)=s_{0}\left(\frac{a_{0}}{a}\right)^{3}.\end{flalign}
Now, assuming that $\rho=\rho\left(T,n\right)$, $p=p\left(T,n\right)$
and using the Gibbs law

\begin{equation}
T\left(\frac{\partial p}{\partial T}\right)_{n}=p+\rho-n\left(\frac{\partial\rho}{\partial n}\right)_{T},\end{equation}
combined with

\begin{equation}
\dot{\rho}=\dot{n}\left(\frac{\partial\rho}{\partial n}\right)_{T}+\dot{T}\left(\frac{\partial\rho}{\partial T}\right)_{n},\end{equation}
and the continuity equation, we obtain\begin{flalign}
 & \left[\left(\frac{3}{2}H+\frac{1}{\left(1+w\left(a\right)\right)}\frac{\dot{R}_{A}}{R_{A}}\right)T\left(a\right)w\left(a\right)+\dot{T}\left(a\right)\right]\left(\frac{\partial\rho}{\partial a}\right)_{n}\nonumber \\
= & -\left(\frac{3}{2}H+\frac{1}{\left(1+w\left(a\right)\right)}\frac{\dot{R}_{A}}{R_{A}}\right)T\left(a\right)\rho\left(a\right)w'\left(a\right).\label{eq:EqDiff}\end{flalign}
Calculating $\left(\frac{\partial\rho}{\partial a}\right)_{n}$ from
Eq.(\ref{eq:rho}), we get the equation governing the evolution of
temperature

\begin{equation}
w'(a)T(a)-\left[\frac{3}{2a}\left(1+w(a)\right)+\frac{R_{A}'}{R_{A}}\right]T(a)w(a)=\left(1+w(a)\right)T'(a).\label{eq:EqDif}\end{equation}
Solving for $w(a)\neq-1,$ we obtain

\begin{equation}
T(a)=T_{0}\left[\frac{w(a)+1}{w_{0}+1}\right]\left[\frac{a_{0}^{3w_{0}/2}}{a^{3w/2}}\right]\exp\left[\int_{a_{0}}^{a}da\left[\frac{3}{2}w'(a)\ln a-\frac{w\left(a\right)}{\left(1+w\left(a\right)\right)}\frac{R_{A}'\left(a\right)}{R_{A}\left(a\right)}\right]\right].\label{eq:Temp}\end{equation}
Using this expression, we write the energy density as a function of
temperature

\begin{equation}
\rho(a)=\rho_{0}\left[\frac{T(a)}{T_{0}}\frac{\left(w_{0}+1\right)}{\left(w(a)+1\right)}\right]\left(\frac{a_{0}}{a}\right)^{\frac{3}{2}}\exp\left[-\int_{a_{0}}^{a}\frac{da}{\left(1+w\left(a\right)\right)}\frac{R_{A}'\left(a\right)}{R_{A}\left(a\right)}\right].\label{eq:Energydensity}\end{equation}
 Extracting the scale factor from (\ref{eq:Temperature-1}), we obtain
the generalized Stefan-Boltzmann law

\begin{flalign}
\rho\left(T\right)= & \rho_{0}\left[\frac{T(a)}{T_{0}}\frac{\left(w_{0}+1\right)}{\left(w(a)+1\right)}\right]^{\frac{w(a)+1}{w\left(a\right)}}a_{0}^{\frac{3}{2}\frac{w(a)-w_{0}}{w\left(a\right)}}\exp\left[-\frac{3}{2w(a)}\int_{a_{0}}^{a}daw'(a)\ln a\right]\nonumber \\
\times & \exp\left\{ \left[\frac{1}{w(a)}\int_{a_{0}}^{a}da\frac{w(a)}{\left(1+w\left(a\right)\right)}-\int_{a_{0}}^{a}\frac{da}{\left(1+w\left(a\right)\right)}\right]\frac{R_{A}'\left(a\right)}{R_{A}\left(a\right)}\right\} .\label{eq:Density}\end{flalign}
 Let us now reproduce the expressions of the temperature and energy
density in the pure spatially flat FRW universe. Indeed, substituting
(\ref{eq:pure}) in (\ref{eq:Density}) and (\ref{eq:Temperature-1}),
we obtain

\begin{equation}
\rho\left(T\right)=\rho_{0}\left[\frac{T(a)}{T_{0}}\frac{\left(w_{0}+1\right)}{\left(w(a)+1\right)}\right]^{\frac{w(a)+1}{w\left(a\right)}}a_{0}^{\frac{3}{2}\frac{w(a)-w_{0}}{w\left(a\right)}}\exp\left[\frac{3}{w(a)}\int_{a}^{a_{0}}daw'(a)\ln a\right]\label{eq:rho0}\end{equation}
 and

\begin{equation}
T(a)=T_{0}\left[\frac{w(a)+1}{w_{0}+1}\right]\left[\frac{a_{0}^{3w_{0}}}{a^{3w}}\right]\exp\left[-3\int_{a}^{a_{0}}daw'(a)\ln a\right],\label{eq:Temp0}\end{equation}
which are exactly the relations obtained in \cite{Saridakis}. 

Let us now scrutinize the behavior of the temperature when $w(a)$
crosses -1. In this case, Eq.(\ref{eq:EqDif}) becomes

\begin{equation}
\left[w'+\frac{R_{A}'(a)}{R(a)}\right]_{\textrm{PDL}}T(a)=0,\label{eq:EqDiff2}\end{equation}
where $\left.w'\right|_{\textrm{PDL}}$ is the value of $w(a)$ at
the phantom divide line (PDL) and $C$ is constant of integration.
The solution of (\ref{eq:EqDiff2}) is:\begin{equation}
R_{\textrm{PDL}}(a)=Ce^{-\left.w'\right|_{\textrm{PDL}}(a-a_{0})},T(a)\neq0,\quad\textrm{or }\quad T(a)=0.\end{equation}
Here, we note that unlike the standard vacuum solution with $T=0$
\cite{Saridakis}, the solution considered in this paper allows for
a vacuum solution with a non-zero temperature. This was expected,
since this temperature is associated with the AH of the black hole
in the absence of dark field. The zero temperature vacuum state is
recovered by setting $M_{0}=0.$ On the other it is important to observe
that the expressions of temperature and energy density are regular
everywhere including the phantom divide crossing. This can be easily
verified by substituting the solution of the AH at the PDL, $R(a)\underset{\textrm{PDL}}{=}Ce^{-\left.w'\right|_{\textrm{PDL}}(a-a_{0})}$
in Eqs.(\ref{eq:Temperature-1},\ref{eq:Energydensity}) \begin{equation}
T_{\textrm{PDL}}(a)=T_{0}a_{0}^{3w_{0}/2}a^{3/2}\exp\left[-1-w_{0}+\frac{3}{2}\left.w'\right|_{\textrm{PDL}}\int_{a_{0}}^{a}da\ln a\right],\label{eq:TempPDL}\end{equation}
 and\begin{equation}
\rho_{\textrm{PDL}}(a)=\rho_{0}a_{0}^{3(w_{0}+1)/2}\exp\left[-1-w_{0}+\frac{3}{2}\left.w'\right|_{\textrm{PDL}}\int_{a_{0}}^{a}da\ln a\right].\end{equation}
 Obviously, Eq.(\ref{eq:TempPDL}) shows that the temperature remains
positive when $w(a)\longrightarrow-1^{\pm}$. Now, making the following
ansatz for the AH \begin{equation}
R_{A}(a)=Ce^{-\left.w'\right|_{\textrm{PDL}}(a-a_{0})}f_{A}(a),\end{equation}
 with $f_{A}(a)\underset{w\rightarrow-1^{\pm}}{\longrightarrow}1$,
we rewrite the temperature as \begin{flalign}
T(a) & =T_{\textrm{PDL}}\left[\frac{w(a)+1}{\left[w(a)+1\right]_{\textrm{PDL}}}\right]a^{-\frac{3}{2}\left(w(a)+1\right)}\nonumber \\
 & \times\exp\left[-\frac{3}{2}\left.w'(a)\right|_{\textrm{PDL}}\int_{a_{0}}^{a}da\ln a+\int_{a_{0}}^{a}da\left[\frac{3}{2}w'(a)\ln a-\frac{w\left(a\right)}{\left(1+w\left(a\right)\right)}\frac{f_{A}'\left(a\right)}{f_{A}\left(a\right)}\right]\right],\label{eq:Temperature}\end{flalign}
which shows that the temperature is always positive definite regardless
the value of $w(a).$ When $M_{0}=0$ the temperature is positive
for $w(a)>1$, negative for $w(a)<-1$ and zero for $w(a)=-1$ \cite{Saridakis}.

Now, we calculate the chemical potential defined by the Euler relation
\cite{Callen}

\begin{equation}
\mu\left(a\right)=\frac{\left(w\left(a\right)+1\right)\rho\left(a\right)}{n\left(a\right)}-\frac{T\left(a\right)s\left(a\right)}{n\left(a\right)}.\label{eq:Def-mu}\end{equation}
Using the expressions of $\rho\left(a\right),$ $n(a),$ $s(a)$ and
$T(a)$ we obtain\begin{equation}
\mu\left(a\right)=\mu_{0}\left[\frac{w(a)+1}{w_{0}+1}\right]\left[\frac{a_{0}^{3w_{0}/2}}{a^{3w(a)/2}}\right]\exp\left\{ \int_{a_{0}}^{a}da\left[\frac{3}{2}w'(a)\ln a-\frac{w\left(a\right)}{\left(1+w\left(a\right)\right)}\frac{R_{A}'\left(a\right)}{R_{A}\left(a\right)}\right]\right\} ,\end{equation}
where

\begin{equation}
\mu_{0}=\frac{\left(w_{0}+1\right)\rho_{0}}{n_{0}}-\frac{T_{0}s_{0}}{n_{0}}\end{equation}
is the present day chemical potential. We note that in general the
sign of $\mu_{0}$ can be arbitrary, and consequently the sign of
$\mu\left(a\right)$.

The entropy of the universe can be derived from (\ref{eq:Def-mu})
and is given by

\begin{equation}
s(a)=\frac{\left(w\left(a\right)+1\right)\rho\left(a\right)-\mu\left(a\right)n\left(a\right)}{T\left(a\right)}.\end{equation}
 Using again the relations for $\rho\left(a\right),$ $T\left(a\right),$
$\mu\left(a\right)$ and $n\left(a\right)$ we obtain

\begin{equation}
s(a)=s_{0}\frac{a_{0}^{\frac{3}{2}}}{a^{\frac{3}{2}}}\exp\left[\int_{a}^{a_{0}}\frac{R_{A}'(a)}{\left(1+w(a)\right)R_{A}(a)}da\right],\label{eq:s-1}\end{equation}
 where the present day entropy is

\begin{equation}
s_{0}=\left[\frac{\left(w_{0}+1\right)\rho_{0}}{T_{0}}-\frac{\mu_{0}n_{0}}{T_{0}}\right].\label{eq:s0}\end{equation}
 Now, defining the comoving volume by

\begin{equation}
V(a)=a^{\frac{3}{2}}\exp\left[\frac{1}{2}\int^{a}\frac{R_{A}'(a)}{\left(1+w(a)\right)R_{A}(a)}da\right],\end{equation}
we derive from Eq.(\ref{eq:s-1}) the usual entropy conservation law

\begin{equation}
s(a)V(a)=s_{0}V_{0}.\end{equation}
As we did for the energy density, we write the entropy in terms of
temperature as

\begin{flalign}
s(T)= & s_{0}\left[\frac{T(a)}{T_{0}}\frac{w_{0}+1}{w(a)+1}\right]^{\frac{1}{w\left(a\right)}}a_{0}^{\frac{3}{2}\left(\frac{w(a)-w_{0}}{w(a)}\right)}\exp\left[-\frac{3}{2w(a)}\int_{a_{0}}^{a}daw'(a)\ln a\right]\nonumber \\
\times & \exp\left[\left[\frac{1}{w(a)}\int_{a_{0}}^{a}da\frac{w(a)}{\left(1+w\left(a\right)\right)}-\int_{a_{0}}^{a}\frac{da}{\left(1+w\left(a\right)\right)}\right]\frac{R_{A}'\left(a\right)}{R_{A}\left(a\right)}\right].\label{eq:Entropy}\end{flalign}
Let us now consider the expressions of the chemical potential and
entropy at the phantom divide crossing. Using (3.21) we easily show
that \begin{equation}
\mu_{\textrm{PDL}}\left(a\right)=\mu_{0}a_{0}^{3w_{0}/2}a^{3/2}\exp\left\{ -1-w_{0}+\frac{3}{2}\left.w'\right|_{\textrm{PDL}}\int_{a_{0}}^{a}da\ln a\right\} ,\label{eq:muPDL}\end{equation}
 and\begin{equation}
s_{\textrm{PDL}}\left(a\right)=s_{0}\frac{a_{0}^{\frac{3}{2}}}{a^{\frac{3}{2}}}\exp\left\{ -1-w_{0}+\frac{3}{2}\left.w'\right|_{\textrm{PDL}}\int_{a_{0}}^{a}da\ln a\right\} .\label{eq:sPDL}\end{equation}
Note that using relations (\ref{eq:TempPDL}), (\ref{eq:muPDL}) and
(\ref{eq:sPDL}) we easily verify that the Euler relation remains
valid at the phantom divide line.

Finally, let us list the corresponding relations for the energy density,
temperature, entropy and chemical potential in the case $w\left(a\right)=w=const,$
\begin{equation}
T(a)=T_{0}\left[\frac{a_{0}}{a}\right]^{3w/2}\left[\frac{R_{A}\left(a_{0}\right)}{R_{A}\left(a\right)}\right]^{\frac{w}{1+w}},\end{equation}
 \begin{equation}
\rho\left(a\right)=\rho_{0}\left[\frac{a_{0}}{a}\right]^{\frac{3}{2}\left(1+w\right)}\left[\frac{R_{A}\left(a_{0}\right)}{R_{A}\left(a\right)}\right],\label{eq:rho-w}\end{equation}
 \begin{equation}
\mu\left(a\right)=\mu_{0}\left[\frac{a_{0}}{a}\right]^{3w/2}\left[\frac{R_{A}\left(a_{0}\right)}{R_{A}\left(a\right)}\right]^{\frac{w}{1+w}},\end{equation}
 \begin{equation}
s(a)=s_{0}\left[\frac{a_{0}}{a}\right]^{\frac{3}{2}}\left[\frac{R_{A}(a_{0})}{R_{A}(a)}\right]^{\frac{1}{1+w}},\end{equation}
 \begin{equation}
\rho\left(T\right)=\rho_{0}\left[\frac{T\left(a\right)}{T_{0}}\right]^{\frac{1+w}{w}},\end{equation}
 \begin{equation}
\mu\left(T\right)=\mu_{0}\left(\frac{T\left(a\right)}{T_{0}}\right),\end{equation}
 \begin{equation}
s(T)=s_{0}\left[\frac{T(a)}{T_{0}}\right]^{\frac{1}{w}}=s_{0}\left[\frac{\rho}{\rho_{0}}\right]^{\frac{1}{1+w}}.\label{eq:s-rho}\end{equation}
It is interesting to note that the dependence of the thermodynamics
parameters on the temperature is of the same form as in standard thermodynamics
of purely expanding FRW universe. 

Before ending this section, let us reconsider explicitly the avoidance
of the big rip singularity in a phantom dominated universe. Using
Eqs.(\ref{eq:continuity},\ref{eq:rho-w}) we obtain\begin{flalign}
a(t) & =a_{0}\left[\frac{3}{2}H_{0}\left(1+w\right)\left(t-t_{s}\right)\right]^{\frac{2}{3\left(1+w\right)}},\label{eq:scalefactor}\\
H(t) & =\frac{2}{3\left(1+w\right)\left(t-t_{s}\right)},\end{flalign}
where the big rip time is \begin{equation}
t_{s}=t_{0}-\frac{2}{3H_{0}\left(1+w\right)}.\end{equation}
Using the constraint (\ref{eq:Hcond}) on $H(t)$, in phantom dominated
era, one finds\begin{equation}
\left(t_{s}-t\right)\geq\frac{16Gm_{H}(t)}{3\left(\left|w\right|-1\right)}.\label{eq:bigRipCond}\end{equation}
Then using the FRW equations and (\ref{eq:GmH}), we find \begin{flalign}
\rho & <\frac{\left(t-t_{s}\right)^{-1}}{8\pi G^{2}m_{H}\left(1+w\right)},\quad\left|p\right|<\frac{w\left(t_{s}-t\right)^{-1}}{8\pi G^{2}m_{H}\left(1+w\right)}.\label{eq:rho-p}\end{flalign}
Substituting (\ref{eq:bigRipCond}) in these relations we reproduce
the relations (\ref{eq:energy-pressure}). Hence, the universe evolves
towards a state where $a,\:\rho,\: p$ and higher derivatives are
finite, without ever reaching the big rip singularity. This final
state is reached in a finite time given by\begin{equation}
t_{*}=t_{s}+\frac{2\left[8GH_{0}M_{0}a_{0}^{\frac{1}{2}\left(5+3w\right)}\right]^{\frac{1+3w}{3\left(1+w\right)}}}{3H_{0}\left(1+w\right)}.\end{equation}
This behavior of the universe is similar to that observed in the framework
of generalized uncertainty principle (GUP) corrected FRW universe
\cite{Nouicer}.

\section{Stability of the solution}

In the following, we focus on the regime of cosmic dynamics where
the universe undergoes a phase of quasi-exponential expansion, such
that $\dot{H}/H^{2}\ll1$, and examine the stability of the solutions
obtained above at the crossing of the phantom divide line when quantum
effects due to conformal anomaly are taken into account. In general,
the conformal anomaly is given by \cite{Duff}\begin{equation}
T=b\left(F+\frac{2}{3}\square R\right)+b'G+b''\square R,\end{equation}
where $F=C_{\mu\nu\lambda\kappa}C^{\mu\nu\lambda\kappa}=\frac{1}{3}R^{2}-2R_{\mu\nu}R^{\mu\nu}+R_{\mu\nu\lambda\kappa}R^{\mu\nu\lambda\kappa}$
is the square of the Weyl tensor and $G$ the Gauss-Bonnet invariant,
$G=R^{2}-4R_{\mu\nu}R^{\mu\nu}+R_{\mu\nu\lambda\kappa}R^{\mu\nu\lambda\kappa}$,
and the coefficient $b$ and $b'$ are\begin{flalign}
b & =\frac{1}{120\left(4\pi\right)^{2}}\left(n_{0}+6n_{1/2}+12n_{1}\right),\\
b' & =\frac{1}{360\left(4\pi\right)^{2}}\left(n_{0}+11n_{1/2}+62n_{1}\right),\end{flalign}
where $n_{0},$ $n_{1/2}$ and $n_{1}$ are the number of scalar,
Dirac fermion and vector fields, respectively. 

Since we have the condition (\ref{eq:Hcond}), we just give the expressions
of $\square R$, $F$ and $G$ to leading order in $Gm_{H}H$ as \begin{flalign}
F & \sim O\left(\left(Gm_{H}H\right)^{2}\right),\quad\square R\sim O\left(\left(Gm_{H}H\right)^{2}\right),\\
G & \simeq42b'H^{4}+128b'H^{5}Gm_{H}.\end{flalign}
Now assuming that\begin{equation}
T_{a}=-\rho_{a}+3p_{a}.\end{equation}
and using the conservation law\begin{equation}
\dot{\rho}_{a}+3H\left(\rho_{a}+p_{a}\right)=0,\end{equation}
one finds \begin{equation}
4\rho_{a}+\frac{1}{H}\frac{d\rho_{a}}{dt}=-T_{a}.\end{equation}
Then $\rho_{a}$ and $p_{a}$ are given by\begin{equation}
\rho_{a}=-\frac{1}{a^{4}}\int dta^{4}HT_{a},\quad p_{a}=\frac{T_{a}+\rho_{a}}{3}.\end{equation}
In the quasi-exponential expansion we can write $\rho_{a}\sim p_{a}\sim T_{a},$
and then we have \begin{equation}
\rho_{a}\sim p_{a}\sim42b'H^{4}+128b'H^{5}Gm_{H}.\end{equation}
On the other hand the fluid density energy can be approximated by\begin{equation}
\rho_{f}\sim p_{f}\sim\frac{3H^{2}}{8\pi G}+\frac{3m_{H}H^{3}}{4\pi G}.\end{equation}
Taking the present day values of the Hubble parameter, $H_{0}\sim10^{-33}$
eV, and $G\sim10^{-56}$ $\left(\textrm{eV}\right)^{-2}$ , it is
easy to verify that the quantum correction are very small when crossing
the phantom divide line. Even we consider BH masses at the end of
the expansion of the order of $\sim10^{23}M_{\odot},$ the correction
term to the standard result is insignificant. Hence, the solutions
given in section 4 are stable under the conformal anomaly quantum
corrections.

\section{Accretion of phantom fluid and constraints on the GSL}

We now consider the problem of the validity of the GSL when the effect
of the back-reaction effect of the phantom fluid on the black hole
is taken into account. We restrict our study to the scenario of a
black hole with a small quasi-local mass immersed in phantom fluid
dominated era. In this case only the cosmological AH contributes,
and the total entropy consists essentially of the sum of entropy of
the cosmological AH and the entropy of the phantom fluid in thermal
equilibrium with the cosmological AH. Indeed, using (\ref{eq:s-rho})
we have

\begin{equation}
S=\left[\frac{\pi R_{C}^{2}}{G}+s_{0}V\left(\frac{\rho}{\rho_{0}}\right)^{\frac{1}{1+w}}\right].\end{equation}
 The first term is the entropy of the cosmological AH and the second
term is the phantom fluid entropy inside a comoving volume $V.$ Taking
the derivative with respect to time and using the following approximation

\begin{equation}
R_{C}\thickapprox\frac{1}{H}-\frac{2G\dot{m}_{h}}{H},\end{equation}
 along with the relations $\ddot{m}_{h}=m_{h}\ddot{a}/a$, $\ddot{a}=a\left(H^{2}+\dot{H}\right)$,
we obtain to leading order in the Hawking-Hayward quasi-local mass\begin{alignat}{1}
\dot{S}= & -\frac{2\pi}{GH}\left[\frac{\dot{H}}{H^{2}}+2G\dot{m}_{H}\left(1-\frac{\dot{H}}{H^{2}}\right)\right]\label{eq:Spoint}\\
 & +\frac{s_{0}VH}{\left(1+w\right)}\left(\frac{T}{T_{0}}\right)^{\frac{1}{w}}\left[\frac{\dot{H}}{H^{2}}+2G\dot{m}_{H}\left(1+\dot{\frac{H}{H^{2}}}\right)-\frac{3}{2}\left(1+w\right)\right],\nonumber \end{alignat}
 where we have used the continuity equation (\ref{eq:continuity})
and (\ref{eq:s-rho}). In order to satisfy the GSL, $\dot{S}\geq0$,
the quantity inside the square brackets must be positive definite,
yielding to the condition

\begin{equation}
2\dot{m}_{H}\geq\frac{\frac{2\pi\dot{H}}{GH^{2}}-\frac{s_{0}VH^{2}}{\left(1+w\right)}\left(\frac{\rho}{\rho_{0}}\right)^{\frac{1}{1+w}}\left[\frac{\dot{H}}{H^{2}}-\frac{3}{2}\left(1+w\right)\right]}{\frac{s_{0}VH^{2}}{\left(1+w\right)}\left(\frac{\rho}{\rho_{0}}\right)^{\frac{1}{1+w}}G\left(1+\dot{\frac{H}{H^{2}}}\right)-2\pi\left(1-\frac{\dot{H}}{H^{2}}\right)}.\end{equation}
 Now substituting $\frac{\dot{H}}{H^{2}}=-\frac{3}{2}\left(1+w\right)$,
we rewrite the constraint on the GSL as

\begin{equation}
G\dot{m}_{H}\geq-3\frac{\left(1+w\right)}{\left(1+3w\right)}\left[\frac{\mathcal{A}-\frac{1+w}{2}}{\mathcal{A}+\frac{\left(5+3w\right)\left(1+w\right)}{1+3w}}\right],\label{eq:GSL}\end{equation}
 with $\mathcal{A=}\frac{s_{0}VH^{2}G}{2\pi}\left(\frac{T}{T_{0}}\right)^{\frac{1}{w}}$.

We know from Eq.(\ref{eq:rate}) that the Hawking-Hayward quasi-local
mass is an increasing function of time in an expanding universe, $\dot{m}_{H}\geq0$.
Then, imposing the positiveness of the RHS of Eq.(\ref{eq:GSL}) and
assuming the positiveness of the entropy, we have one solution given
by

\begin{equation}
-\frac{(5+3w)\left(1+w\right)}{\left(1+3w\right)}>\mathcal{A}\geq0,\label{eq:phantom}\end{equation}
which is satisfied for an EoS parameter in the range $w\leq-5/3.$
Using the expression of $s_{0}$ given in (\ref{eq:s0}) and defining
$\alpha_{0}=-\mu_{0}n_{0}/\rho_{0},$ an immediate consequence of
the positiveness of the entropy is that $w\geq-1-\alpha_{0}.$ Since
$w\leq-5/3$, we get a lower bound estimate to the present day value
of the chemical potential, \begin{equation}
\alpha_{0}\geq\frac{2}{3},\label{eq:b-alpha}\end{equation}
 which show that the chemical potential in phantom energy dominated-era
is strictly negative. Now, with $w=-2$ and the present day parameters,
we obtain the following bound on the entropy\begin{equation}
s_{0}<3.5\times10^{-4}\,\textrm{GeV}^{-3}.\label{eq:bound-s}\end{equation}
 Or in terms of $\alpha_{0}$ and using (\ref{eq:b-alpha}), we have\begin{equation}
\frac{2}{3}\leq\alpha_{0}<1+1.26\times10^{43}T_{0}.\label{eq:bound-alpha}\end{equation}
 Now, if we take $T_{0}\backsimeq10^{-19}$ GeV we have $\alpha_{0}\lesssim10^{24}.$
Here we would like to point that the bounds on the present day values
of the entropy and the parameter $\alpha_{0}$ are of the same order
of that obtained by Lima et al. \cite{Lima-Horvath}. However, the
calculation in this paper is performed by ignoring the back-reaction
of the phantom fluid on the black hole. In fact, they used the Schwarzschild
metric, which cannot describe the properties of black holes embedded
in an expanding FRW universe. As we explicitly show, taking into account
the back-reaction effect, even in a low density background, leads
to a drastic constraint on the EoS parameter and rule out the approach
based on zero chemical potential \cite{Horvath}.

Now, using the obvious relation $m_{H}=\dot{m}_{H}/H,$ we obtain
the quasi-local black hole mass above which the accretion of phantom
fluid onto the black hole is permitted\begin{equation}
m_{H}\geq-\frac{3}{GH}\frac{\left(1+w\right)}{\left(1+3w\right)}\left[\frac{\mathcal{A}-\frac{1+w}{2}}{\mathcal{A}+\frac{\left(5+3w\right)\left(1+w\right)}{1+3w}}\right].\end{equation}
 Using the constraint on the GSL (\ref{eq:phantom}), we obtain\begin{equation}
m_{H}\gtrsim m_{H,crit}=-\frac{6}{GH}\left[\frac{\mathcal{A}-\frac{1+w}{2}}{\left(5+3w\right)}\right].\end{equation}
 In terms of $\alpha_{0}$ and the present day quantities, the critical
quasi-local mass is then\begin{equation}
m_{H,crit}=-42.27\times10^{-20}\left[\frac{1+w+\alpha_{0}}{5+3w}\right]\frac{M_{\odot}}{T_{0}}.\end{equation}
For large values of $\alpha_{0}\thicksim10^{24}$ and $T_{0}\thickapprox10^{-19}$
GeV, the critical quasi-local mass is approximately $m_{H,crit}\thickapprox10^{23}M_{\odot}.$
This is a huge value, allowing all black holes in the universe to
accrete phantom fluid. We note that in section 3, we have shown that
$m_{H,crit}$ is the maximal allowed mass which prevents the engulfing
of the universe by the BH. On the other hand small values of $\alpha_{0}$
give critical BH mass of the order of the solar mass. For instance,
taking $\alpha_{0}=2$, $w=-2$ and $T_{0}\thickapprox10^{-19}$ GeV,
we obtain $m_{H,crit}\thickapprox4.2M_{\odot}.$ Our results are similar
to that obtained in \cite{Lima-Horvath}, where the EoS parameter
is restricted to values less than $-1$ and the back-reaction effect
ignored. We have plotted the present day critical quasi-local mass
with respect to $w$ for different values of the parameter $\alpha_{0}.$
Obviously, the value of the parameter $\alpha_{0}$ is very important
in determining a critical quasi-local mass of the order of the solar
mass. Finally, if we set the chemical potential to zero, the critical
mass reduces to \begin{equation}
m_{H,crit}=-\frac{3}{2\pi}\frac{\left(1+w\right)}{\left(5+3w\right)}\frac{\rho_{0}VH}{T_{0}}\left(\frac{T}{T_{0}}\right)^{\frac{1}{w}},\end{equation}
which is negative for $w<-5/3.$ This result is expected from the
relation (\ref{eq:b-alpha}).

\begin{figure}[H]
\begin{centering}
\includegraphics[width=8cm,height=8cm]{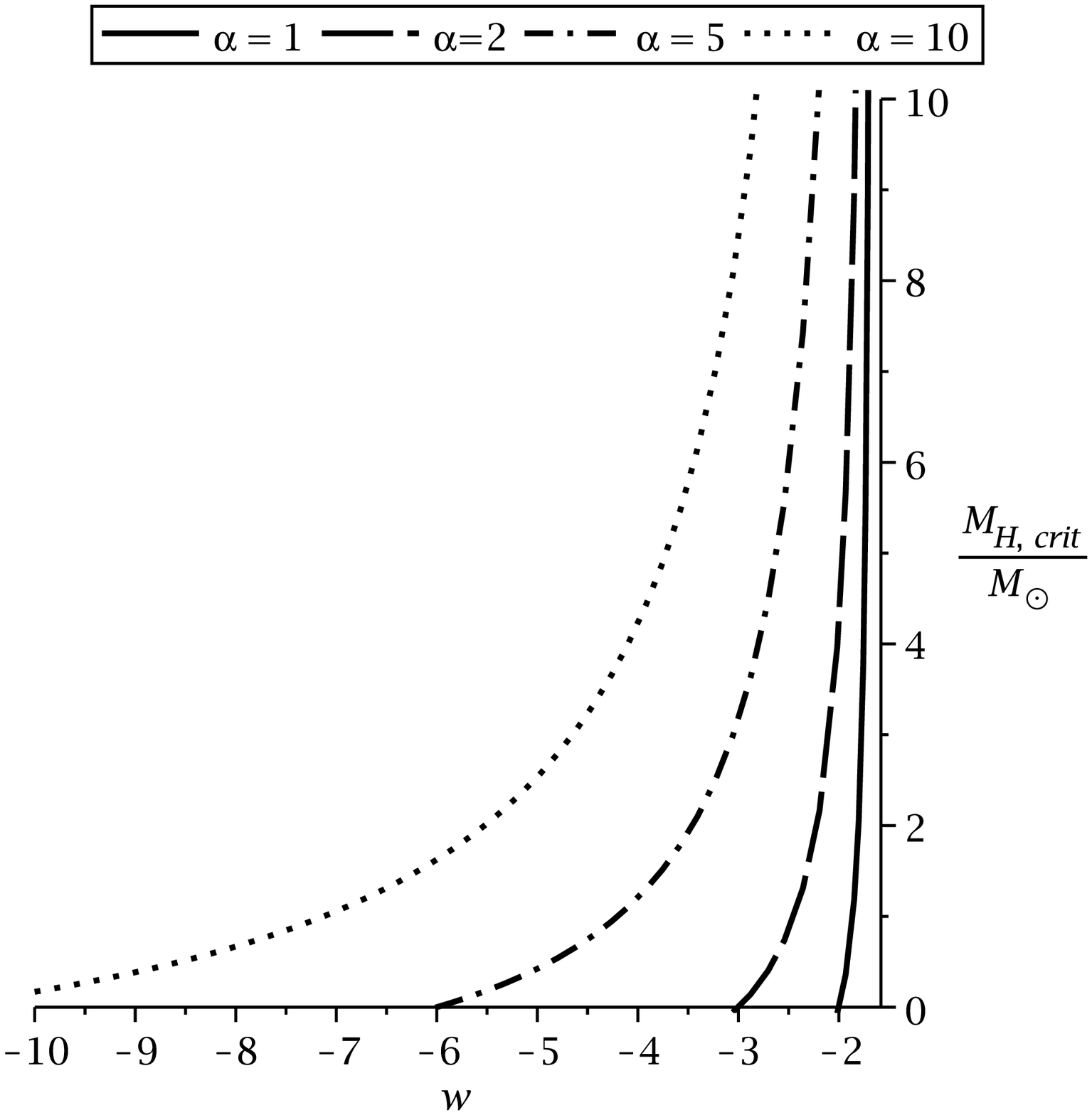} 
\par\end{centering}

Figure 3: Variation of the critical Hawking-Hayward quasi-local mass
with respect to $w$ for different values of the parameter $\alpha_{0}.$ 
\end{figure}

\section{Conclusion}

In summary, we have investigated the thermodynamical properties of
black holes immersed in an expanding spatially flat FRW universe,
for general EoS parameter $w\left(a\right)$. Particularly, we found
that the temperature of the dark fluid is always positive regardless
the value of the time varying EoS parameter, and that the instantaneous
vacuum state is characterized by non-zero temperature, entropy and
chemical potential, respectively. An other important result is that
all the thermodynamics parameters are regular at the phantom divide
crossing. We have also analyzed the accretion process of phantom fluid
onto black holes with small Hawking-Hayward quasi-local mass, and
the constraints on the validity of the GSL. Particularly, assuming
the positiveness of the phantom fluid entropy, we have shown that
phantom fluid may have a negative definite chemical potential in order
to satisfy the GSL. We have also obtained, for an EoS parameter within
the interval $w<-5/3,$ a critical quasi-local mass of the black hole,
above which the GSL is always protected. The present analysis, show
that taking into account the back-reaction effect of the phantom fluid
on the black hole, even in a low density background, leads naturally
to positive temperature and negative chemical potential, and may contributes
to resolve the controversy on the subject \cite{Horvath,Lima-Horvath}.

\section*{Acknowledgments}

The author was supported by the Algerian Ministry of High Education
and Scientific Research under the CNEPRU project N\textdegree{}:
D01720070033.

\end{document}